\documentclass[aps,prd,showpacs,nofootinbib,superscriptaddress,preprint,tightenlines]{revtex4}
\pdfoutput=1
\usepackage{graphicx}
\usepackage{amsbsy}

\usepackage{amsfonts}

\usepackage{amssymb}

\usepackage{amsmath}
\newcommand{\hateq}{\widehat{=}}
\def\be{\nopagebreak[3]\begin{equation}}
\def\ee{\end{equation}}
\def\ba{\nopagebreak[3]\begin{eqnarray}}
\def\ea{\end{eqnarray}}

\def\d{{\rm d}}

\def\U(1){{\rm U(1)}}

\def\d{{\rm d}}

\def\bar{\overline}

\def\={\hateq}
\def\SU(2){\rm SU(2)}
\def\U(1){\rm U(1)}

\newcommand{\teta}{\rlap{\lower2ex\hbox{$\,\tilde{}$}}\eta{}}

\begin{document}
\preprint{\vbox{\baselineskip=12pt \rightline{UNAM-IM-MOR-2009-1}
}}
\title{Black holes and entropy in loop quantum gravity: An overview}
\author{Alejandro Corichi}
\email{corichi@matmor.unam.mx} \affiliation{Instituto de
Matem\'aticas, Unidad Morelia, Universidad Nacional Aut\'onoma de
M\'exico, UNAM-Campus Morelia, A. Postal 61-3, Morelia,
Michoac\'an 58090, Mexico}
\affiliation{Center for Fundamental Theory,
Institute for Gravitation and the Cosmos,
Pennsylvania State University, University Park
PA 16802, USA}

%\maketitle

\begin{abstract}
Black holes in equilibrium and the counting of their entropy
within Loop Quantum Gravity are reviewed. In particular, we focus
on the conceptual setting of the formalism, briefly summarizing
the main results of the classical formalism and its quantization.
We then focus on recent results for small, Planck scale, black
holes, where new structures have been shown to arise, in
particular an effective \emph{quantization} of the entropy. We
discuss recent results that employ in a very effective manner
results from number theory, providing a complete solution to the
counting of black hole entropy. We end with some comments on other
approaches that are motivated by loop quantum gravity.
\end{abstract}
 \maketitle
%\draft
%\clearpage%

\section{Introduction}
%\vskip0.25cm

Black holes (BH) have become rather prominent in fundamental
physics ever since the fundamental results in the early 70's
showing that black holes satisfy some `thermodynamic-like laws',
summarized in the celebrated laws of black hole mechanics
\cite{BCH},
\be \delta M=\frac{\kappa}{8\pi G}\; \delta A\, ,\label{BHMec}
\ee
From which one can formally relate,
\[
 M\leftrightarrow E, \qquad
\kappa \leftrightarrow T\, , \qquad A \leftrightarrow S\, ,
\]
where the relation between {\it geometrical variables} on
Eq.~(\ref{BHMec}) can be seen as the analogue of the first law of
thermodynamics if the above association between geometric and
thermodynamical objects is made. This analogy is further motivated
by the fact that the surface gravity $\kappa$ of a Killing horizon
is constant and the area of an event horizon always grows. This
observation, together with the proposal by Bekenstein  and Hawking
that BH possess a physical entropy and temperature, as confirmed
by the computation of particle creation on black hole background,
gave raise to a true identification between geometrical quantities
and thermodynamical variables as follows \cite{BH}:
\[ E=M\qquad
T=\frac{\kappa\;\hbar}{2\pi}\quad
{\rm and}\quad
S=\frac{A}{4\,G\hbar}\, .
\]
It is not unnatural to interpret that black holes must behave as
thermodynamic systems, and in particular possess a non-zero
temperature (that vanishes in the classical limit) and an entropy
(that blows up). Quantum theory was needed in order to identify
temperature and entropy with geometrical objects, by means of
Planck's constant $\hbar$, suggesting that these identifications
are quantum in nature. But, in order to have a full analogy, the
question of what are the underlying degrees of freedom responsible
for entropy became a pressing one. In other words, how can we
account for the (huge) entropy associated to the black hole
horizons?

The standard wisdom is that only with a full marriage of Gravity
and the Quantum will we be able to understand this issue. This is
one of the main challenges that faces any candidate quantum theory
of gravity.

During the past 20 years there have been several attempts to
identify those degrees of freedom. In particular one has to
mention the success of string theory in explaining the entropy of
extremal and near-extremal BH in several dimensions~\cite{vafa}.
There have also been some proposals based on causal
sets~\cite{causal} and on the use of entanglement entropy of
matter fields~\cite{entanglement}. Within loop quantum
gravity~\cite{lqg,lqg:easy}, a leading candidate for a quantum
theory of gravity, there has been some progress in describing
black holes `in equilibrium'. In particular this implies that the
objects to be studied are assumed to be isolated, in such a way
that a study of its properties will guarantee that one can
separate their description from that of the rest of the
environment (as one normally does in thermodynamics). The
resulting quantum picture is that the interaction between `bulk
states' as described by {\it spin networks} as they puncture the
horizon, create horizon degrees of freedom that can (and do)
fluctuate. These degrees of freedom are, on the one hand,
independent of the bulk degrees of freedom, and on the other hand,
fluctuate `in tandem' with their bulk counterparts, as dictated by
specific quantum conditions warranting the existence of the
\emph{quantum horizon}.

The original program was developed in a series of
papers~\cite{ABCK, ACK, ABK} and has been further
studied~\cite{AEV, Dom:Lew, majumdar, meiss, GM, CDF, CDF-2, ADF,
ABDFV} and also widely reviewed in~\cite{LR,CDF-3}. The purpose of
this contribution is to provide a bird's eye view into the field,
briefly summarizing the progress made in the past 12 years,
including some recent results. This contribution can also be seen
as a starting point and as a reading guide for those interested in
more details.

In what follows, we shall in particular try to answer the
following questions: How do we characterize black holes in
equilibrium? That is, what are the quantum horizon states? How do
we know which states we should count? Can we learn how entropy
behaves? Can we make contact, for large black holes, with the
Bekenstein-Hawking entropy? Can we extend the formalism and
consider small, Planck scale BH's? How small is small? That is,
where does the transition from the Planck scale to the `large area
limit' occurs?

We shall not include topics such as the possibility of treating
Hawking radiation~\cite{carlo,kirill} or the criteria for dealing
with black holes in thermal equilibrium~\cite{majumdar2}.
%A summary of complementary material not covered here can be found, for instance, in Ref.~\cite{majumdar3}.

\section{Preliminaries}
%\vskip0.25cm

This section has two parts. In the first one we review the
motivation for the need of a notion of horizon that is local and
not teleological as is the case of the traditional event horizon.
In the second part we briefly review the main ideas behind the
isolated horizons formalism.

\subsection{Motivation}

\noindent Physically, one is interested in describing black holes
in equilibrium. That is, equilibrium of the horizon, not the
exterior. Just as in the standard analysis of physical systems
subject to thermodynamics considerations, one requires {\it the
system} and not the whole universe to be in equilibrium. The use
of globally stationary solutions to Einstein's equations to study
the thermodynamics of horizons is very restrictive since one is
requiring the whole universe to be stationary and not just the
system, i.e. the horizon. Can one capture that notion via
quasi-local boundary conditions? Yes! And the answer is provided
by the isolated horizons (IH) formalism~\cite{ACK,LR}.

The main idea is that \emph{some} boundary conditions are imposed
on an inner boundary of the spacetime region under consideration.
The interior region of the horizon is cut out, since the isolated
horizon is regarded as a boundary. Is this a physical boundary?
No! but one can ask whether one can make sense of it, namely
whether there is a consistent prescription for incorporating this
hypothesis, and a consistent variational principle is possible. A
second question pertains to the physical interpretation of the
boundary. If `physics' does not end there, in the sense that in a
realistic spacetime, matter and observers can fall into the
interior region with a well defined evolution, what is then the
justification for `arbitrarily' cutting this region out?

The justification is that, being null surfaces, the exterior
region (say in an asymptotic region) will not have access to any
events inside the horizon (even if the isolated horizon does not
coincide with a possible event horizon, it will lie {\it inside
it}), the information of what happens inside is not needed for
describing the physical processes in the outside region. One can
then interpret the horizon, and the degrees of freedom on it, as a
`screen' that keeps track of those aspects of the degrees of
freedom that fell in but that can still interact with the outside
region. For instance, the mass of the isolated horizon has certain
information of the energy of the matter that fell in, and is
responsible for the gravitational field outside the horizon. The
same is true for other quantities such as charge, angular
momentum, etc. These horizon charges (multipole moments) will
carry this information, and is the input needed in formulating the
theory.

Let us summarize the main features of IH and their quantum treatment:\\

i) The boundary $\Delta$, the 3-D isolated horizon, provides an effective description
of the degrees of freedom of the {\it inside region}, that is cut out in the formalism.\\

ii) The boundary conditions are such that they capture the intuitive description of a
horizon in classical equilibrium and allow for a consistent variational principle.\\

iii) The quantum geometry of the horizon has independent degrees
of freedom that fluctuate `in tandem' with the bulk quantum geometry. \\

iv) The quantum boundary degrees of freedom are then responsible for the
entropy. \\

v) The entropy thus found can be interpreted as the entropy assigned by an `outside
observer' to the (2-dim) horizon $S=\Sigma \cap \Delta$.

\vskip0.15cm

\begin{centerline}{
\includegraphics[angle=270,scale=0.50]{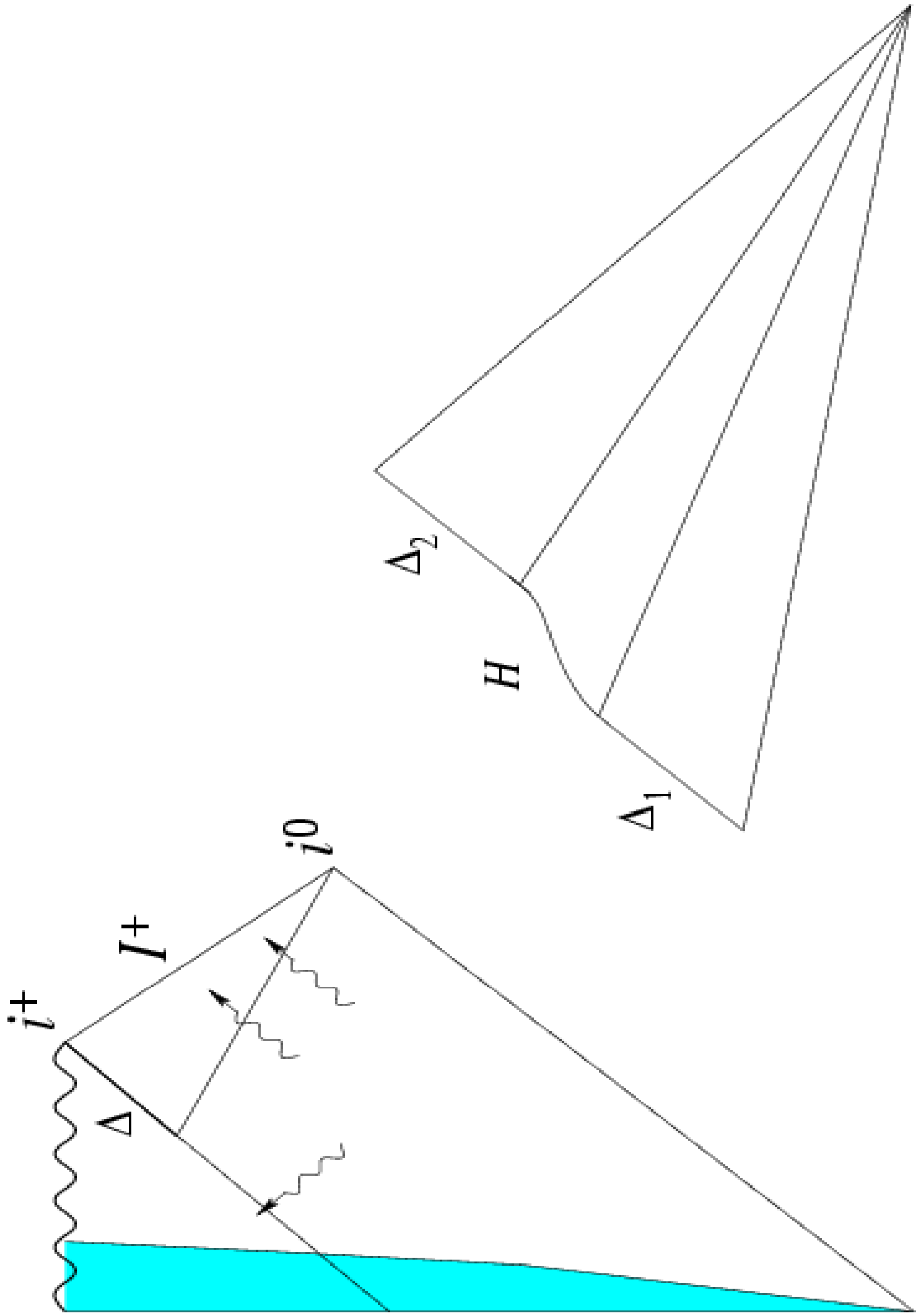}
} \vskip0.15cm {\noindent Fig.1 Left: The physical situation one
expects to describe. The collapse of a stellar object creates an
event horizon that settles down (rather quickly) and in the
asymptotic future is non-expanding, giving rise to an Isolated
Horizon $\Delta$. Right: even if there is more matter falling in
the future, there will be portions of the horizon that will be
isolated.}
\end{centerline}

\vskip0.15cm

\noindent Just as for other approaches to black hole entropy, the
LQG treatment is not free from some interpretational issues. For
instance, is the entropy to be regarded as the entropy contained
by the horizon? Is there some `holographic principle' in action?
Can the result be associated to entanglement entropy between the
interior and the exterior?, etc. Some of these questions have been
clarified but there are still some for which we have no answer yet
(see for instance the discussion in \cite{MRT}).

\subsection{Isolated Horizons}
%\vskip0.15cm

In this part we will provide the main ideas in the definition of
isolated horizons. For full details see \cite{LR}. An isolated
horizon is a null, non-expanding 3D-surface $\Delta$, equipped
with some notion of translational symmetry along its generators
(it is assumed to have a congruence of null vectors generating
it). There are three main consequences of these boundary
conditions:

\vskip0.15cm \noindent i) The gravitational degrees of freedom
induced on the horizon are captured by a $U(1)$ connection, \be
W_a = -\,\frac{1}{2}\; \Gamma^i_a\,r_i \ee where $\Gamma^i_a$ is
the spin connection of the canonical theory on $\Sigma$. Thus,
there is an effective reduction of the gauge symmetry from $SU(2)$
to $U(1)$.

\vskip0.15cm \noindent ii) The total symplectic structure of the
theory (and this is true even when matter is present) gets split
as, \be
 \Omega_{\rm tot}=\Omega_{\rm bulk} +\Omega_{\rm hor}
\ee
with
$${\Omega_{\rm hor}=\frac{a_0}{8\pi\,G}\oint_S{\rm d}W\wedge{\rm d}W^\prime}$$
This is precisely the symplectic structure one would get if we
were considering a Chern-Simons theory for the U(1) connection
$W_a$ on the three dimensional manifold $\Delta$ with $S$  a
spatial section (Recall that Chern Simons does not require a
metric, so the fact that $\Delta$ is null is irrelevant).

\vskip0.15cm \noindent iii) Finally, the `connection part' and the
`triad part' at the horizon must satisfy the condition, \be
\label{hor-cons}
F_{ab}=-\,\frac{2\pi\,\gamma}{a_0}\,E^i_{ab}\,r_i\, , \ee the so
called `horizon constraint'. Here $F_{ab}$ is the curvature of the
U(1) connection $W_a$.

\subsection{Constraints}
%\vskip1.5cm

\noindent It is interesting to explore the consequences of the
boundary conditions in the Hamiltonian framework. A detailed study
of the canonical theory \cite{ACK,LR} reveals an interesting
structure. In particular, the formalism tells us what is gauge and
what not. To be precise, with respect to the constraints that
appear in the canonical formalism, we know that:

\vskip0.15cm a) The relation between curvature and triad, the
horizon constraint (\ref{hor-cons}),
is equivalent to Gauss' law.%\\

\vskip0.15cm b) Diffeomorphisms that leave $S$ invariant (i.e.
that, when restricted to the horizon map $S$ to itself) are gauge
(i.e. the vector fields generating
infinitesimal diffeomorphisms are tangent to $S$).%\\

\vskip0.15cm c) The scalar constraint must have a vanishing lapse
$N|_{\rm hor}=0$ at the horizon. Thus, the gauge transformations
generated by the scalar constraint (that depend of the lapse),
leave the horizon untouched. In particular, this implies that any
gauge and diff-invariant observable {\it is} a full Dirac
observable. This list includes all multipole moments of the
horizon.

This last point is the reason behind the fact that one can
sensibly talk about the quantum theory of black holes in LQG even
when we have not solved the quantum dynamics in the bulk. That is,
since in the quantum theory one has to implement the constraints,
the fact that the lapse vanishes at the horizon implies that, from
the horizon perspective, any quantum state that satisfies Gauss'
law and is diffeomorphism invariant will by a physical state,
given that the Hamiltonian constraint imposes no further
condition. Of course, one has to make sure that the quantum
horizon states `interact' properly with the bulk states for which
the dynamics is still not fully understood. This represents one of
the current challenges.

\section{Quantum Theory: The Bulk}
%\vskip0.25cm

\noindent
Loop quantum gravity \cite{lqg} is based on a canonical
formulation of general relativity in terms of connections and
triads (For a brief introduction see \cite{lqg:easy}).

The basic canonical variables are: \be A^i_a     \quad {\rm
a}\;SU(2)\;{\rm connection}\quad ;\quad E^a_i \quad{\rm a\;
densitized\; triad} \ee with $A^i_a=\Gamma^i_a-{ \gamma}\,K^i_a$,
and $\gamma$ real the Barbero-Immirzi parameter (BI). Loop Quantum
gravity defined on a manifold without boundary is based on two
fundamental observables of the basic variables:
\be
{\rm Holonomies}, \quad h_e(A):={\cal P}\exp(\int_e A)
\ee
and
\be {\rm Electric\; Fluxes}, \quad E(f,S):=\int_S \d S^{ab}
E_{i\,ab}\,f^i\, . \ee
where $E^c_i=\tilde{\eta}^{abc}\,E^i_{ab}$. The main assumption of
Loop Quantum Gravity is that these quantities become well defined
operators in the quantum theory. Thus, the starting point for LQG
is the so called Holonomy-Flux algebra ${\cal HF}$ \cite{ACZ}. An
important question is how many consistent representations of the
${\cal HF}$-algebra there are. In recent years, the LOST
collaboration proved the following result: There is a unique
representation of the Holonomy-Flux algebra on a Hilbert space
that is {\it diffeomorphism invariant} \cite{LOST}. This
representation corresponds precisely to the construction of
Ashtekar and Lewandowski \cite{lqg}. Let us now give a brief
description of this resulting Hilbert space. First we can
characterize it in terms of Spin Networks: \be {\cal H}_{\rm AL}=
\oplus_{\rm graphs}\,{\cal H}_\Upsilon={\rm Span\; of\;all\; Spin
\;Networks}\;{|\Upsilon, \vec{j}, \vec{m}\rangle} \ee \vskip0.3cm

\begin{centerline}{
\includegraphics[angle=270,scale=.50]{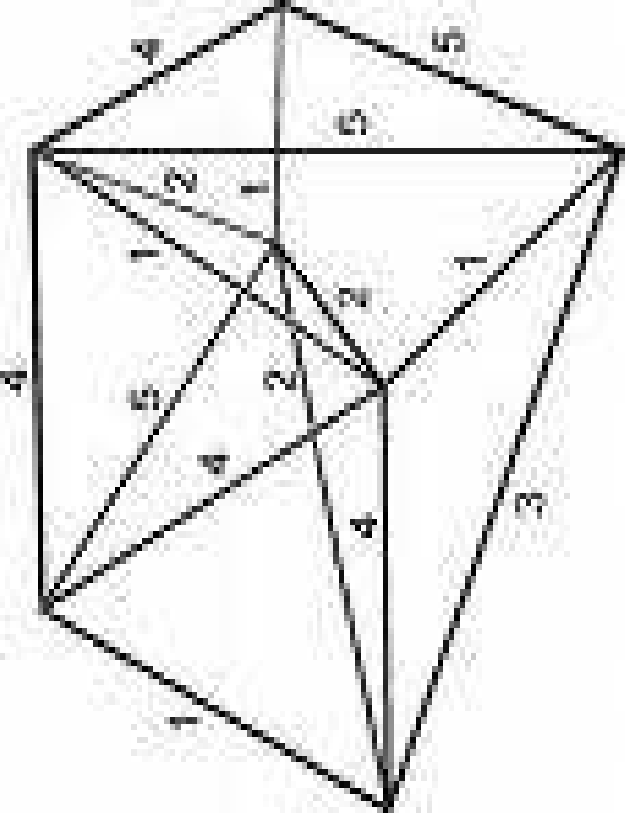}
} {\noindent Fig 2. A Spin network $(\Upsilon,\vec{j},\vec{m})$
consists of a graph $\Upsilon$ together with labels $j_i$ for the
edges and $m_i$ for the vertices.}
\end{centerline}

\vskip0.5cm \noindent A Spin Network {$|\Upsilon, \vec{j},
\vec{m}\rangle$}, represents a particularly convenient basis for
the theory. It is a state labelled by a graph $\Upsilon$, and some
colorings $(\vec{j},\vec{m})$ associated to edges and vertices.

The spin networks have a very nice interpretation in terms of the
quantum geometry they generate. They are the eigenstates of the
quantized geometry, such as the area operator,
\be
\hat{A}[S]\cdot{|\Upsilon, \vec{j},\vec{m}\rangle}=8\pi
\ell_{\rm Pl}^2{ \gamma}\sum_{\rm edges}
\sqrt{j_I(j_I+2)}\;{|\Upsilon, \vec{j},\vec{m}\rangle}
\ee
where the sum is over all the intersection points $p_I$ of the
edges $e_I$ with the surface $S$. The standard interpretation is
that the edges of the graph excite the quantum geometry of the
surface $S$ at the intersection points between $S$ and $\Upsilon$.
The edges $e_I$ can be seen as quantum fluxes of area.

\begin{centerline}{
\includegraphics[angle=270,scale=.65]{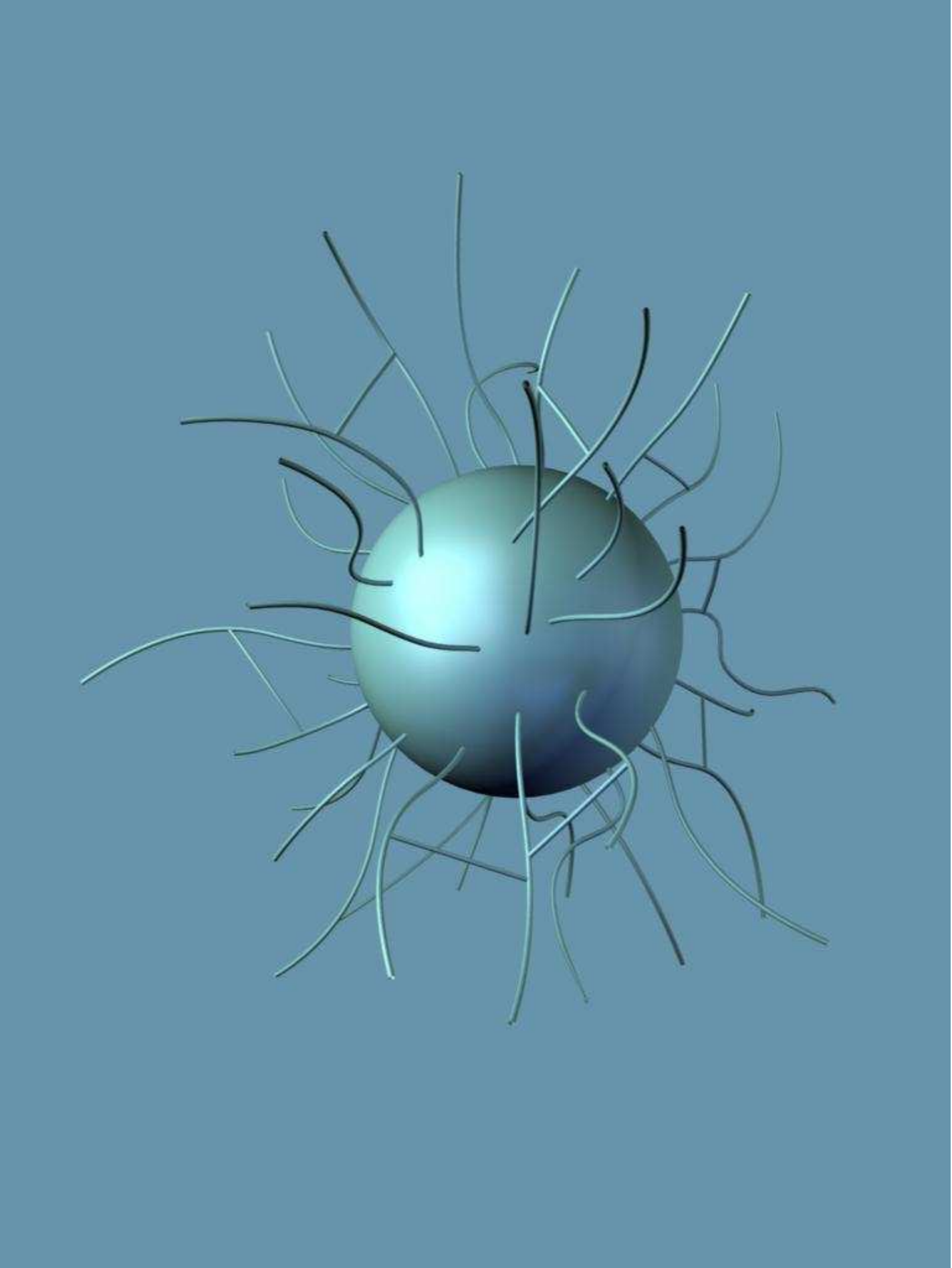}}
\vskip0.4cm {\noindent Fig.3. An artist impression of a black hole
in LQC. The edges of the state on the bulk puncture the horizon
$S=\Sigma \cap \Delta$ endowing it with area through the labels
$j$'s and with intrinsic curvature through the $m$'s. }
\end{centerline}

\section{Quantum Theory of the Horizon}
%\vskip0.25cm

\noindent Just as in the classical description of the
gravitational field with an IH, the phase space could be
decomposed in a bulk part and a horizon part, a basic assumption
is that the total Hilbert Space is a tensor product of the form:
\be {\cal H} = {\cal H}_{\rm V} \otimes {\cal H}_{\rm S} \ee where
${\cal H}_{\rm S}$, the surface Hilbert Space, can be built from
Chern Simons Hilbert spaces for a sphere with punctures. This
represents the `kinematical Hilbert space'.

In order to go to the physical theory, the conditions on ${\cal
H}$ that we need to impose are: Invariance under diffeomorphisms
of $S$ and the quantum condition on $\Psi$, the quantum
equivalence of Eq.~(\ref{hor-cons}): \be \left({\rm Id}\otimes
\hat{F}_{ab}+\,\frac{2\pi\,\gamma}{a_0}\,\hat{E}^i_{ab}\,r_i
\otimes {\rm Id} \right)\cdot \Psi=0\, .\label{q-cond} \ee Then,
the theory we are considering is a quantum gravity theory, with an
isolated horizon of fixed area $a_0$ (and multiple-moments). A
Physical state would be such that, in the bulk satisfy the
ordinary constraints and, at the horizon, the {\it quantum horizon
condition}. \vskip.15cm

\noindent{\it Entropy.} We shall consider the simplest case of
pure gravity with a non-rotating horizon. In this case, from the
outset, we are given a black hole of area $a_0$. The question then
is: what entropy can we assign to it? Let us take the
microcanonical viewpoint. To compute the entropy we shall count
the number of states ${\cal N}$ such that they satisfy:
\vskip.15cm

\noindent
$\bullet$ The area eigenvalue is in the interval
$\langle \hat{A}\rangle\in [a_0-\delta, a_0+\delta]$\\
%\vskip1.5cm

\noindent
$\bullet$ The quantum horizon condition (\ref{q-cond}) is satisfied.
\vskip.15cm

\noindent The entropy ${\cal S}$ will be then given by, \be {\cal
S}=\ln{\cal N}\, .
\ee
The challenge now is to identify those states that satisfy the two
conditions, and count them. \vskip.15cm

\noindent{\it Characterization of the States.} There is a
convenient way of characterizing the states by means of the spin
network basis. If an edge of a spin network with label $j_I$ ends
at the horizon $S$, it creates a puncture, with label $j_I$. The
area of the horizon will be the area that the operator on the bulk
assigns to it: $A=8\pi{\gamma} \ell^2_{\rm
Pl}\sum_i\sqrt{j_I(j_I+1)}$.

Is there any other quantum number associated to the punctures
$p_I$? Yes! They are given by eigenstates of $\hat{E}_{ab}$ that
are also half integers $m_I$, such that $- j_I\leq m_I\leq j_I$.
The quantum horizon condition relates these eigenstates to those
of the Chern-Simons theory. The requirement that the horizon is a
(topological) sphere then imposes a `total projection condition'
on $m's$:
\be
  \sum_I m_I=0
\ee
that has to be taken into account as well.

A quantum horizon state can be conveniently characterized by a set
of punctures $p_I$ and to each one a pair of half integer
{$(j_I,m_I)$}, where the three previous conditions impose some
restrictions on the possible values of the labels.

If we are given $N$ punctures and two assignments of labels
$(j_I,m_I)$ and $(j'_I,m'_I)$. Are they physically
distinguishable? or a there some `permutations' of the labels that
give indistinguishable states? That is, what is the statistics of
the punctures?

As usual, we should let the theory tell us. One does {\it not} postulate any statistics.
If one treats in a careful way the action of the diffeomorphisms on the
punctures one learns that when one has a pair of punctures with the same labels
$j$'s and $m$'s, then the punctures are indistinguishable and one should not count them
twice. In all other cases the states are distinguishable.
\vskip.15cm

\noindent{\it The counting.} We start with an isolated horizon,
with area $a_0$ (assumed to be of the order of several Planck
areas) and ask how many states are there compatible with the two
conditions, and taking into account the distinguishability of the
states. One can approach the problem in a two step process.

\noindent First step: Count just the different configurations and
forget about $\sum_I m_I=0$. Thus, given $\{n_j\}^{j_{\rm
max}}_{j=1/2}=(n_{1/2},n_1,n_{3/2},\ldots,n_{s_{\rm max}/2})$,
where $n_j$ means the number of punctures with label $j$,we count
the number of states:
\be
{\cal N} = \frac{N!}{\Pi_j\, (n_{j}!)}\Pi_j\,(2j + 1)^{n_{j}}
\ee
with $N=\sum_j n_j$. Taking the {\it large area approximation}
$A>> \ell_{\rm Pl}$, and using the Sterling approximation, one
gets as the dominant term: \be S=\frac{A}{4\ell^2_{\rm Pl}}
\;\frac{\gamma_0}{{\gamma}} \ee with $\gamma_0$ the solution%
\footnote{This counting was first done in detail in \cite{GM}.
There is a slightly different counting (sometimes denoted as the
DLM counting) that does not distinguish configurations with
different $j$'s if the $m$'s are the same. In that case we get a
different linear dependence with $\sum_j
2\,e^{2\pi\,\gamma_M\sqrt{j_i(j_i+1)}}=1$ \cite{Dom:Lew,meiss}.)}
to $\sum_j (2j+1)e^{2\pi\,\gamma_0\sqrt{j_i(j_i+1)}}=1$.

As a second step one introduces the projection constraint. This
has the effect of introducing a correction to the entropy area
relation as an infinite series, where the first correction is
logarithmic \cite{majumdar,meiss,GM}: \be S=\frac{A}{4\ell^2_{\rm
Pl}} \;\frac{\gamma_0}{{\gamma}} - {\frac{1}{2}}\ln(A) + \ldots
\ee Note that one gets, in the complete counting, the asymptotic
linear dependence on area. If we want to make contact with the
Bekenstein-Hawking formula we have to make use of the freedom in
LQG provided by the BI parameter and choose
$\gamma=\gamma_0$.\footnote{As noted before, the value of
$\gamma_0$ depends on the counting} The coefficient of the
logarithmic correction seems to be universal and independent of
the particular counting (for other topologies of the horizon, it
might change \cite{topo}). An important observation is that the
formalism can be generalized to more general situations, the
combinatorial problem is the same and therefore the result is that
{\it the same value of $\gamma$ will yield the BH entropy}. These
more general horizons include arbitrary distortion and rotation in
vacuum gravity \cite{AEV} as well as coupling to electromagnetic,
dilatonic, Yang-Mills, cosmological constant \cite{ACK,ABK}, and
non-minimally coupled scalar fields \cite{AC}.

 In the following sections we shall review new
developments that have occurred since 2006. These include a new
phenomena found when computing directly the number of states for
small black holes, and an exact counting of states by use of
methods from number theory.

\section{Direct Countings and Entropy Quantization}
%\vskip0.25cm

In this section we will describe the results found when
considering small Planck size horizons for which the counting of
states is possible. For that one tells a computer how to count for
a range of area $a_0$ at the Planck scale \cite{CDF}. With the
availability of having an exact algorithm under control, one can
ask, for instance, what is the effect of considering or not the
`projection constraint'. In the large area approximation it is
responsible for the first, logarithmic, correction term. One could
also ask when is the linear dependence with area first observed.
That is, when do we see a transition from deep  quantum effects to
`large areas'? Let us briefly summarize the results reported in
\cite{CDF}. What was found is that, without the projection
constraint, the entropy approaches very fast a `smooth' function
of area with the slope found in the analytical calculations. When
including the constraint, the relation between entropy and area
became oscillatory, with a well identified period $\delta A_o$,
that {\it on average}, introduced the expected logarithmic
corrections. This was already identified for horizons that are as
small as 100 $\ell^2_{\rm Pl}$. For details see \cite{CDF}.

%\begin{figure}
 % \begin{center}
  %\rotatebox{angle=90}{
 % \includegraphics[angle=0,scale=.75]{espectroln}
  %  \bigskip
%\caption{\label{fig:1} The entropy as a function of area is shown,
%where the projection constraint has not been imposed. The BI is
%taken as $\gamma=0.274$.%}
  %\end{center}
%\end{figure}

%

% \includegraphics[angle=0,scale=.75]{ampliacionln}

%

%\section{ENTROPY QUANTIZATION}

Furthermore, it was seen that, by analyzing the `black hole
spectrum' (i.e. the degeneracy of states as function of area),
both the oscillations found with a large value of $\delta$ as well
as these structures in the `spectrum' possess the same periodicity
$\delta A_0 \approx 2.41\;\ell^2_{\rm Pl}$. A natural question is
whether there is any physical significance to this periodicity. It
turns out that, if one chooses the interval: $2\,\delta = \Delta
A_0$, the plot of the entropy {\it vs} area becomes
stair-like\cite{CDF-2}, as can be seen in Fig.4:

\vskip0.25cm

\begin{centerline}{
\includegraphics[angle=0,scale=.50]{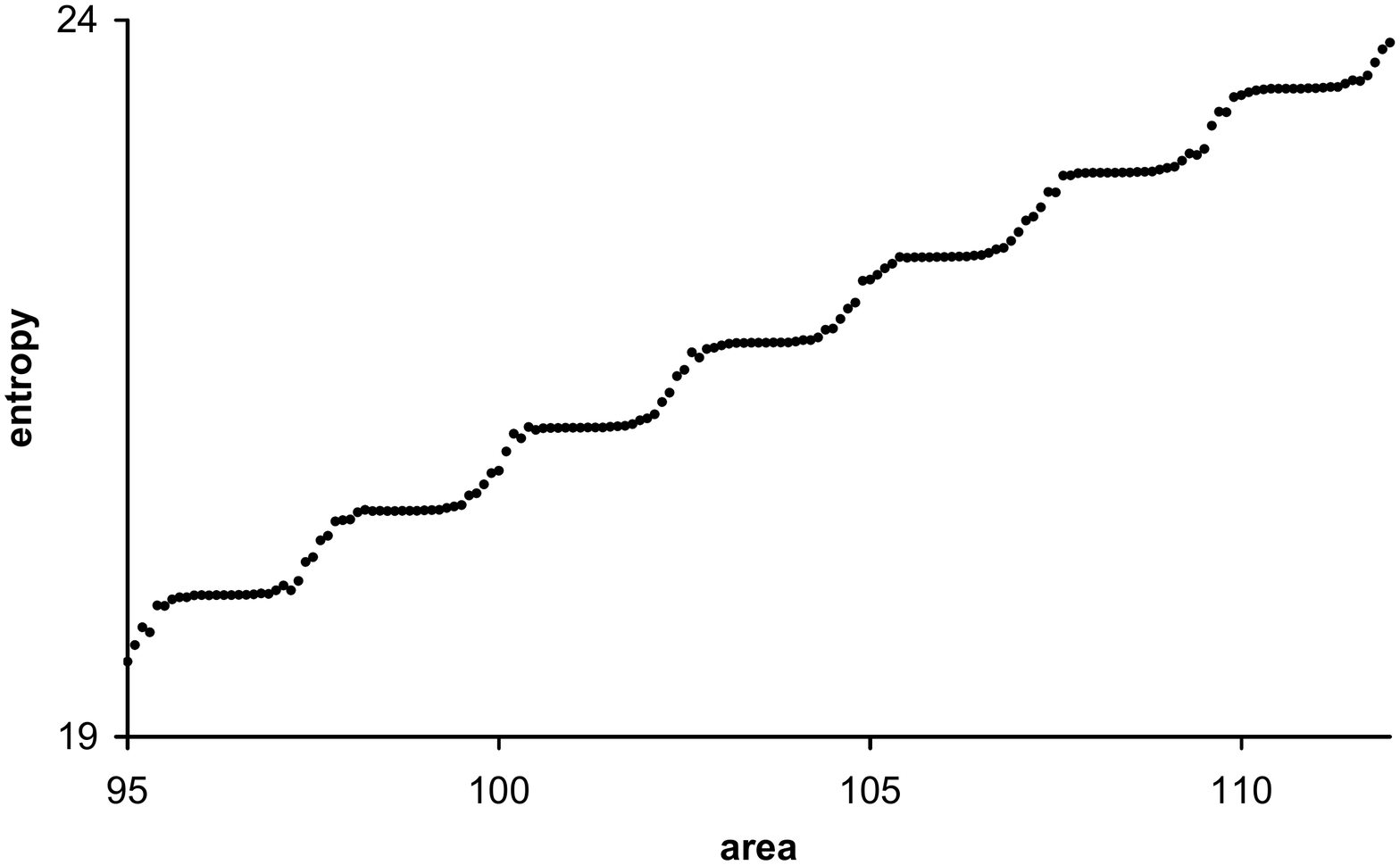}
}
{\noindent Fig. 4. The entropy as function of area shows a step-like behavior when the
interval $\delta$ is chosen to coincide with the periodicity.}
\end{centerline}

\vskip0.3cm

What one notes is that the entropy has a completely different
behavior for this particular choice of interval: Instead of
oscillations, the entropy seems to increase in discrete steps.
Furthermore, the height of the steps seems to approach a constant
value as the area of the horizon grows, thus implementing in a
rather subtle way the conjecture by Bekenstein that entropy should
be equidistant for large black holes. Quite remarkably, this
result is robust, namely, it is {\it independent} of the counting.

While the constant number in which the entropy of large black holes `jumps' seems to
approach \cite{CDF-2}:
\be
\Delta S\mapsto 2\,\gamma_0\,\ln{(3)}
\ee

\begin{centerline}{
\includegraphics[angle=0,scale=.50]{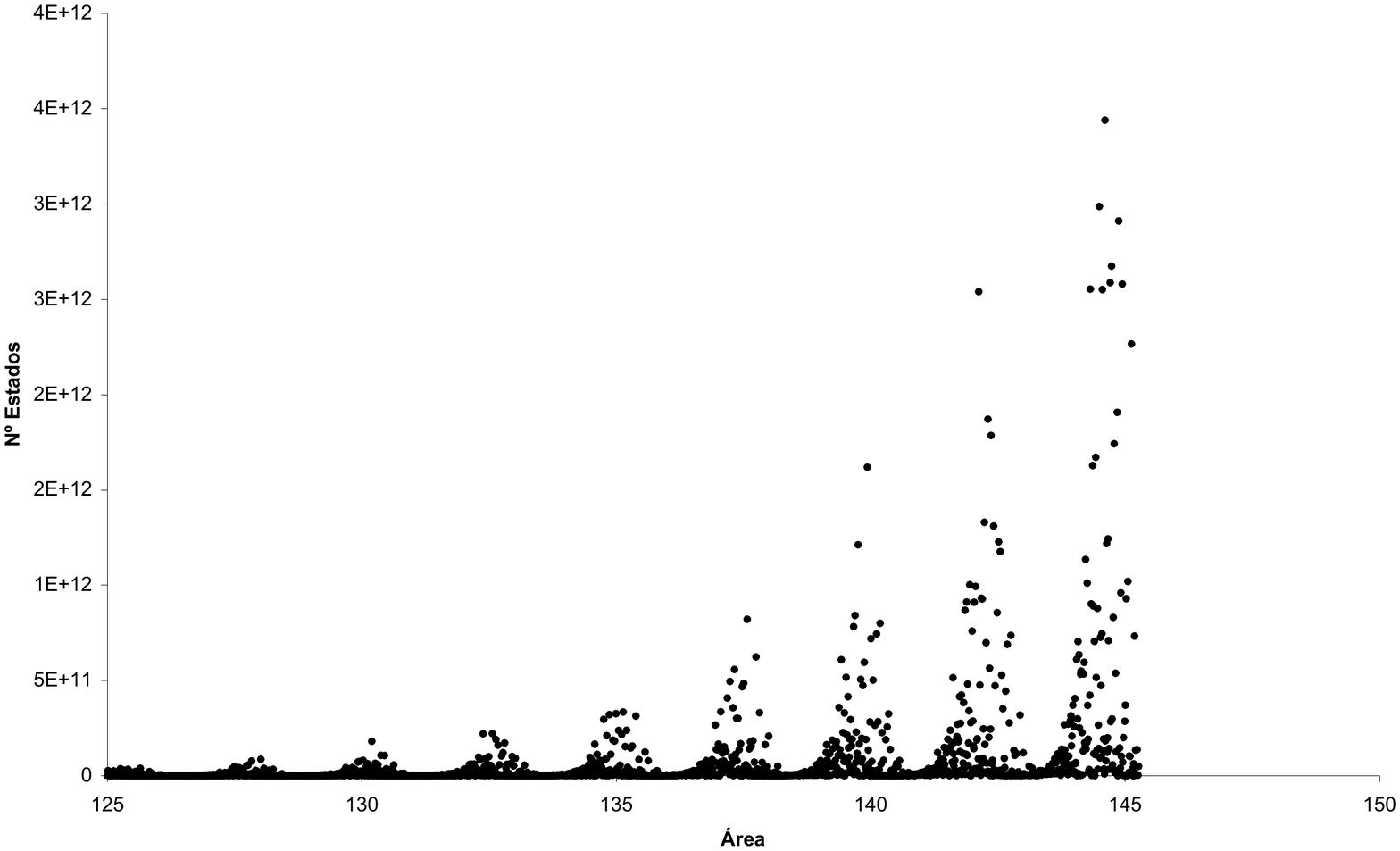}
} {\noindent Fig. 5. The black hole spectrum shows some peaks of
higher degeneracy together with some valleys. This is the origin
of the step-like behavior of entropy.}
\end{centerline}

\vskip0.3cm

\noindent Some recent proposals have provided a heuristic
understanding of the origin of these peaks and valleys
\cite{hanno,ADF}. To summarize these results, the model there
proposed allows one to think of the states as organized in {\it
bands}, labelled by certain combination of the total number of
punctures and `spin'. By employing the analytic $n_j$ distribution
that maximizes degeneracy, as originally introduced in \cite{GM},
one can find the `average area', for each band associated with
this maximum degeneracy configuration, from which one can compute
the change in area from peak to peak as,
\be \Delta A
=\frac{8\pi\gamma\sum_s\sqrt{s(s+2)}(s+1)e^{-2\pi\gamma_0\sqrt{s(s+2)}}}{
3\,\sum_s s(s+1)e^{-2\pi\gamma_0\sqrt{s(s+2)}}+2} \ee
An interesting observation is that if one parametrizes this number
as $\Delta A=\chi\gamma$, one can see that $\chi$ must be a
constant, independent of the counting (since it only depends on
the degeneracy of the states as functions of $j$'s and $m$'s).
From the observed periodicity in the direct counting, it was
conjectured in \cite{CDF-2} that the value of $\chi$ is $8\ln{3}$.
Interestingly, the approximate formula found in \cite{hanno,ADF}
for both countings, yield a slightly different approximate values
$\bar{\chi}$ for the parameter $\chi$, with $\bar{\chi}_{\rm
DLM}<8\ln{3}<\bar{\chi}_{\rm GM}$, and the relative difference of
the order of $10^{-4}$. This shows that the approximation is not
exact and one needs a better analytical understanding of the
combinatorial problem.

\section{Exact Counting: Number Theory}

Recent progress using number theoretical considerations has turned
out  to be useful for the purpose of understanding the emergence
of the discrete structures \cite{ABDFV}. In this study, a
reformulation of the counting of states and an exact counting of
the number of states has been achieved recently. In this part I
shall briefly summarize these results. There are two main steps
involved in the counting. In the first one, one finds a complete
characterization of the area spectrum, that is, of the eigenvalues
of the area operator in terms of  so-called `square-free numbers'.
Then, given an allowed area-eigenvalue, one computes the number of
possible `sets' of labels (be them $j's$ and/or $m's$) that are
compatible with that values. This sets contain also the number of
punctures that have a given label assigned. In the second part of
the counting process one assigns a degeneracy to each `set' coming
from the possible `permutation' of labels. At the end, one obtains
an exact number of consistent states for the given value of area.
The final step, that is, the computation of the entropy can then
be computed either by considering an interval as previously
defined, or by summing over all values of area up until the
prescribed value $A_0$.

Let us now describe how one achieves the first step in the
counting process. First, one notes that the area eigenvalues (when
measured in units of $8\pi G\gamma$ can be written as:
\be A=\sum_{I=1}^{N}\sqrt{(k_I+1)^2-1}=\sum_{k=1}^{k_{\rm
max}}n_k\sqrt{(k+1)^2-1}\label{ec-chida}
\ee
where $k_I=2J_I$ are integers labelling the punctures and we have
recast the sum by rearranging the punctures by their label $k$
($n_k$ is the number of punctures with label $k$) and summing over
labels. The idea here is to employ the square free numbers as a
basis for the area eigenvalues (the numbers are `linearly
independent' under arbitrary linear combinations with integer
coefficients). Each of the terms in the sum can be recast as an
integer $q_i$ times a `square free number' $\sqrt{p_i}$ (by means
of the prime decomposition of the quantities inside the square
root). Thus the sum becomes $\sum_{i=1}^{r}q_i\sqrt{p_i}$, where
$\sqrt{(k+1)^2-1}=y\sqrt{p_i}$,  for some integer $y$.

Let us summarize. If we specify a square free number $p_i$, we
want to know for which values of integers $k$ and $y$ is the
previous equation satisfied, which would tell us (for each
possible solution) the allowed values of the labels $k$. This
equation is known as the Pell equation and has an infinite number
of solutions (labelled by $m$). We can then use these solutions to
rewrite (\ref{ec-chida}) as
\[
A=\sum_{i=1}^{r}\sum_{m=1}^{\infty} n_{k^i_m}y^i_m\,\sqrt{p_i}
\]
If we use the fact that the numbers $\sqrt{p_i}$ are linearly
independent, we can split the equation in a series of different
equations of the form $\sum_{m=1}^{\infty} n_{k^i_m}y^i_m=q_i$,
where the $y$'s and the $q$'s are known, as solutions to the Pell
equation, and the unknowns here are the numbers $n_{k^i_m}$. If
these Diophantine equations admit solutions $\sum_{i=1}^r
q_i\sqrt{p_i}$, then $A$ belongs to the relevant part  of the
spectrum of the area operator, the numbers $k$'s give the spins
involved, and the numbers $n$'s count the number of times that
edges labelled by the spin $k^i_m/2$ pierce the horizon.

Thus, given a linear combination of square free number as the area
eigenvalue, the procedure here described provides an answer to the
`degeneracy' associated to the different  pairs $\{k^i_m,
n_{k^i_m}\}$ defining the different {\it spin configurations}. The
next step in order to obtain the total number of states is to
count the `$m$-degeneracy', namely the different ways of
accommodating the $m$'s on a given spin configuration. It is at
this point that the two different countings (GM and DLM) provide
different answers. Both cases can be treated in terms of {\it
fusion numbers} and fusion matrices employed in CFT. For details
see \cite{ABDFV}. Just as an illustration, for the DLM counting
the answer can be exactly written as
\[
\frac{2^N}{M}\sum_{s=0}^{M-1}
    \prod_{I=1}^{N}\cos(2\pi sK_I/M)
\]
with $M=1+\sum_{I=1}^N k_I$, allowing to have exact expressions
for the degeneracy of states. Of course, this strategy confirms
the results of \cite{CDF-2}, but also allows to compute the
spectrum of larger black holes with the same computational
capacity. These results represent a starting point for more
refined asymptotic analysis, by means of generating functions for
the combinatorial problem, that will shed more light on the
behavior of macroscopic black holes \cite{fernando}. For instance,
an important question to be addressed is whether the oscillatory
behavior on entropy, the  \emph{entropy quantization}, together
with its possible implications for Hawking radiation, is still
present for large black holes.

%\subsection{Discussion}

\section{Other Approaches}

Let us now discuss some open questions regarding quantum black
holes and the progress that has been made within LQG. By the mere
fact that in the isolated horizon framework one is considering
only the outside region of a spacetime containing a back hole, one
is not addressing the issue of the singularity. The possible
singularity resolution has been analyzed in a series of papers
using loop quantization techniques \cite{bhinlqc}.

The starting point of such treatments is the minisuperspace of
homogeneous cosmologies on a spatial manifold with topology
$S^2\times\mathbb{R}$. These `Kantowski-Sachs models' are
important given that the interior region of the global
Schwarzschild solution belongs to this class. It is thus natural
to attempt to employ the same techniques that have been extremely
useful in the treatment of (minisuperspaces corresponding to)
homogeneous and isotropic models in cosmology (See, for instance
\cite{CS:unique} for a recent summary of such methods).

Those results suggest that the classical singularity inside the
horizon, just as in the case of isotropic cosmologies, is avoided,
and the quantum evolution continues past it, but more work is
needed to reach a definite conclusion. In particular, none of the
presently available models \cite{bhinlqc} is able to overcome
consistency requirements that select a unique quantization in the
isotropic sector \cite{CS:unique}.

An important open issue is how to specify black hole/horizon
states from the full theory. That is, without assuming that there
was a classical horizon to begin with. Some progress in this
direction has been made in two fronts, but still at some
preliminary stage. A proposal for defining coherent states that
approximate a spacetime with a black hole, have been used to count
the number of black hole states \cite{dasgupta}. Even when
potentially important, this approach is still in its early stages
given that the coherent states are kinematical, and there is not a
full control on the dynamical sector of the theory.

Another proposal for identifying black hole states from the full
set of states was made within the context of symmetry reduced
models in Ref.~\cite{hus:win}. Here, the idea is to specify
operators on a kinematical Hilbert space such that they project
the kinematical states onto a `black hole sector'. Just as in the
previous case, there are still several consistence criteria that
this approach must satisfy before one can make concrete
predictions. In particular both approaches lack a description for
dynamical processes that the horizon might undergo.

If the singularity resolution were also generic, and there existed
a spacetime interpretation beyond the `would be singularity', one
would be lead to the Ashtekar-Bojowald paradigm for evaporation
and (lack of) information loss \cite{paradigm}. This picture
includes the description of dynamical processes that are no longer
described by the Isolated Horizon formalism. One needs then to
consider the more general framework of {\it dynamical horizons}
\cite{LR}. If this picture is physically correct, there is no
classical singularity and no event horizon ever forms. Still,
there {\it is} a horizon that forms, grows and then shrinks due to
Hawking radiation. Information is not lost, even when, for certain
observers, Hawking radiation appears to be thermal (for more
details within a simple model see \cite{CGHS}). These results are
certainly intriguing and suggest a resolution of the `information
loss problem' due to true quantum geometric effects. Needless to
say, more work is needed to unravel this mystery.

\section{Conclusions and Outlook}
%\vskip0.25cm

Let us summarize what we have learned from the merger of isolated
horizons and loop quantum gravity. First, as we have shown,
isolated horizons provide a consistent framework to incorporate
black holes that are physically in equilibrium, as classical
objects. As we have argued, one can consistently quantize the
theory, as described by the IH phase space, employing both the
methods of quantum geometry that are useful in the bulk, together with
techniques from $U(1)$ Chern-Simons theory on a sphere.
%This provides a full characterization of the theory.
A detailed study of the action of the constraints allows us to
give a full characterization of the  quantum horizon degrees of
freedom that contribute to the entropy. It is found that the
entropy is {\it finite}, without the need of a regulator nor a
cut-off, and that its dominant term is linear in area for large
horizons in Planck units. Furthermore, the formalism allows us to
translate the entropy counting into a purely combinatorial problem
for which one can attempt algorithmic {\it brute force}
computations \cite{CDF}, as well as number-theoretic treatments
\cite{ABDFV}.

A very important feature of this formalism is that one can
incorporate and count the entropy of a whole class of different
black holes, where one can include arbitrary distortion and
rotation in vacuum gravity \cite{AEV} as well as coupling to
electromagnetic, dilatonic, Yang-Mills \cite{ACK,ABK} and
non-minimally coupled scalar fields \cite{AC}. In all these cases
the combinatorial problem to be solved is {\it the same} (even
when its translation into physically relevant quantities might
vary) and therefore, entropy is {\it always} proportional to area
in the large area limit (or with the expected contribution from
the scalar field in the non-minimally coupled case).

As we have explored, when one considers the problem of a direct
counting of the number of states, and thus being forced to
consider small horizons, several unexpected features appear for
these Planck size black holes. While one recovers the asymptotic
linear dependence on area and the logarithmic correction (with the
right coefficient), from which we can say something about BI
parameter, a new behavior is observed for small horizons. It is
found that there are oscillations in entropy with a constant
periodicity. Furthermore, when properly interpreted, this points
to an effective {\it quantization} of the entropy in equidistant
steps \cite{CDF-2}. This observed behavior suggests that loop
quantum gravity can make contact, in a rather subtle manner, with
both Bekenstein's heuristic model \cite{CDF-2}, and the
Mukhanov-Bekenstein effect \cite{Bekenstein,DF}. Whether this
scenario is realized or not remains an intriguing question.

Recently, attempts to understand the origin of the `black hole
spectrum' responsible for the entropy quantization have been put
forward \cite{hanno,ADF,ABDFV}, which have been able to give some
intuitive understanding of the effect. In particular, there has
been some progress to understand, from  a heuristic perspective,
the origin of the `bands' in the spectrum and their equidistant
nature. A pressing question here is whether the discrete
structures found at the Planck scale are still present for
macroscopic black holes. One would also like to understand whether
the constant $\chi$ actually is equal to $8\ln{3}$ as the
numerical computations and the heuristic considerations seem to
suggest. If this were the case, one would need to understand its
origin.

As we have here tried to convey, in the past couple years there
has been exciting progress in our understanding of quantum black
holes within Loop Quantum Gravity, but there are still important
question that remain open regarding the detailed relation between
gravity,  entropy and the quantum.

\section*{Acknowledgments}

\noindent My understanding of black holes in loop quantum gravity
grew from the interaction with numerous colleagues to whom I am
grateful, including A. Ashtekar, J. Baez, J. D\'\i az-Polo, J.
Engle, E. Fern\'andez-Borja, K. Krasnov, J. Lewandowski, C.
Rovelli and H. Sahlmann. I would like to thank J.P. Ruiz-D\'\i az
for Fig.3. This work was in part supported by CONACyT U47857-F
grant, by NSF PHY04-56913 and by the Eberly Research Funds of Penn
State.

\end{document}